\documentclass{article}

\usepackage[english]{babel}

\usepackage[letterpaper,top=2cm,bottom=2cm,left=3cm,right=3cm,marginparwidth=1.75cm]{geometry}

\usepackage{amsmath}
\usepackage{latexsym}
\usepackage{abstract}
\usepackage{cite}
\usepackage{graphicx}
\usepackage{amssymb}
\usepackage{xcolor}
\usepackage[colorlinks=true, allcolors=blue]{hyperref}
\usepackage{ytableau} 
\usepackage{caption} 
\usepackage{subcaption} 

\newcommand{\deltabar}{\delta\hspace*{-0.2em}\bar{}\hspace*{0.1em}}
\newcommand{\dbar}{d\hspace*{-0.08em}\bar{}\hspace*{0.1em}}

\title{\Huge \bf Flavour-kinematics duality \\ for Goldstone modes}
\author{Dijs de Neeling, Diederik Roest and Sam Veldmeijer\footnote{d.w.de.neeling@rug.nl, d.roest@rug.nl, s.r.veldmeijer@student.rug.nl }}
\date{\it Van Swinderen Institute for Particle Physics and Gravity, \\ University of Groningen, the Netherlands
}

\newcommand{\sm}{$\sigma$-model}
\begin{document}
\maketitle
\begin{abstract}
\noindent Three scalar effective field theories have special properties in terms of non-linear symmetries, soft limits and on-shell constructability that arise from their Goldstone nature: the non-linear \sm, multi-DBI theory and the special Galileon. We discuss how these theories are related via flavour-kinematics duality, analogous to the colour-kinematic duality between gravity and gauge theories. At the off-shell level, we identify a specific mapping between the three theories that is crucially dependent on their non-linear symmetries. Similarly, we demonstrate how the on-shell amplitudes factorise into BCJ numerators describing flavour and a scalar version of kinematics, naturally leading to the inclusion of graviton exchange in the $SO(M,N)$ non-linear \sm. Finally, we map those numerators onto each other, and comment on a similar relation to tensor kinematics. Our results highlight a common structure that underlies the physics of different Goldstone modes.
\end{abstract}
\newpage
\tableofcontents

\section{Introduction and outline}

\noindent
General relativity (GR) and Yang-Mills (YM) theory are amongst the central pillars of 20th century physics and describe the gravitational interaction and gauge forces, respectively. Due to different gauge symmetries, spins and quantum (non-)renormalisation properties, these were for a long time thought to have little similarities. However, following the work of \emph{Kawai, Lewellen, and Tye (KLT)} on open-closed string duality, where closed-string amplitudes can be written as a sum over products of open-string amplitudes \cite{Kawai:1986ah}, it has become natural to ask whether (non-)gravitational amplitudes in field theory are similarly related.

A concrete realisation of this duality was proposed by \emph{Bern, Carrasco and Johannsson (BCJ)}, showing that YM and GR tree-level amplitudes can be factorised in a specific way \cite{Bern:2008qj}. The resulting colour and kinematic factors (often referred to as BCJ numerators) can be chosen to satisfy group-theoretical constraints corresponding to Jacobi identities. This algebraic correspondence between kinematics and colour, and hence between  gauge theory and gravity, is widely referred to as \emph{colour-kinematics (CK) duality} (see e.g.~\cite{Elvang, Cheung-TASI, Bern:2019prr} for reviews on amplitudes and CK duality).

Once colour-kinematic duality is satisfied by the \emph{BCJ numerators}, the gravitational tree-level amplitudes can be  written as the ``square'' of their gauge theory counterpart. Following this approach, the YM and GR amplitudes can be written as 
\begin{equation}
  A_n^{\rm YM}=\sum_{i \in \text { cubic }} \frac{C_{i} T_{i}}{D_i}\,,
 \qquad A_n^{\rm GR}=   \sum_{i \in \text { cubic }} \frac{T_{i} T_{i}}{D_i}          \,, \label{factorisation}
\end{equation}
where $C_i$, $T_i$ and $D_i$ respectively denote colour factors, kinematic numerators and propagator structures. The sums run over a set of inequivalent diagrams $i$ that are purely trivalent, i.e.~cubic (and hence differ from the usual Feynman diagrams for these theories). Given a YM amplitude, its gravitational counterpart is constructed by simply interchanging the YM colour factors $C_i$ by another kinematic numerator $T_i$. This procedure is famously referred to as the \emph{double copy}\footnote{Partial progress on the mechanism underlying the double copy was provided by the identification of the kinematic algebra of the self-dual sector of YM theory \cite{Monteiro:2011pc, Chen:2019ywi, Monteiro:2013rya}.} and it has been proven to hold at tree-level \cite{Bridges:2019siz, Bjerrum-Bohr:2010pnr, delaCruz:2017zqr, Mafra:2011kj, Bjerrum-Bohr:2016axv, Du:2017kpo,Carrasco:2020ywq}. Furthermore, CK duality was also conjectured to extend to loop level \cite{Bern:2010ue} and it has been verified in various non-trivial cases \cite{Bern:1998ug, Bern:1997nh, Carrasco:2011mn, Bern:2013yya}. However, a general loop-level proof is still missing.

\bigskip

\noindent
Double copy relations are by no means restricted to GR and YM; a natural third theory, completing the triplet in \eqref{factorisation}, consists of two colour factors and corresponds to cubic interactions of \emph{bi-adjoint scalars} (BAS), in the adjoint of both colour factors. Moreover, similar relations have been found for a large web of different theories, including other spin-one theories as well as scalar effective field theories \cite{Carrasco:2019qwr,   Cheung:2017ems}. The latter include the so-called non-linear (NL) \sm~(whose BCJ factorisation was shown in \cite{Chen}) as well as the special Galileon (SG), and can be built by including a BCJ factor that only depends on momentum and hence describes \emph{scalar-kinematic numerators} (see \cite{Du:2016tbc} for explicit expressions). Similar to the triplet BAS-YM-GR for colour and (tensor-)kinematics, this relates the BAS, NLSM and SG \cite{Cheung-TASI}. 

The NLSM and SG are so-called exceptional scalar field theories, with very special properties in terms of non-linearly realised symmetries and Goldstone mode interpretations. In fact, they can be constructed by specifying the amount of derivatives per field while enforcing an \emph{enhanced soft degree} beyond the Adler zero\footnote{Very recently, non-integer soft degrees have been identified in \cite{Brauner2022}.} \cite{Cheung:2014dqa, Cheung:2016drk}. The latter leads to nontrivial cancellations among Feynman diagrams of different topology, thereby completely constraining the interactions of the theories and their symmetry transformations. Due to this structure, their on-shell structure follows from a soft bootstrap approach: all amplitudes\footnote{See \cite{Joyce} for a very recent extension of soft limits and recursion to the off-shell wavefunction of these theories.} can be seen to follow from a single seed interaction, see e.g.~\cite{Elvang-soft, Low-soft}. Relatedly, the SG satisfies the equivalence principle and can be phrased in terms of diffeomorphisms \cite{Bonifacio, Roest}. These scalar EFT properties clearly echo the corresponding features of gravity and gauge theories \cite{Weinberg, Deser}.

However, these aspects also raise an interesting question, as there is a third theory with such properties, being \emph{Dirac-Born-Infeld} (DBI) theory with multiple scalars \cite{multiDBI}. Inspired by the original colour-kinematics duality relating BAS, YM and GR, is there a similar relation\footnote{These three scalar EFTs also allow for a CHY representation of their amplitudes \cite{Cachazo}.} between NLSM, DBI and SG? We will show that this is indeed the case, and that it corresponds to the introduction of a fourth BCJ numerator, that we refer to as \emph{flavour factors}. In the resulting web of dualities, the NL \sm~then appears twice: either with one colour or with two flavour factors. Are these theories identical? And if not, how do they differ given the strong symmetry constraints on the theory describing pion scattering? The answer to these questions involves graviton exchange in an interesting manner, and is the topic of {\bf section 3}.

In the same section, we will also demonstrate intriguing relations between the different BCJ numerators. Note that this goes beyond the double copy relations of colour-kinematics (or flavour-kinematics) duality: instead of replacing factors to go from one theory to another, these factors themselves also turn out to be related. In particular, we will show how the scalar-kinematic factor follows from the flavour factor (but not vice versa). Although not the focus of this paper, we also propose a non-invertible mapping from tensor-kinematic numerators to the flavour factors\footnote{Putting these together maps tensor-kinematics directly onto scalar-kinematics. This is closely related to the recent results of  \cite{Cheung:2021zvb}, that however uses a trace basis instead of the half-ladder basis that will be central in this article.}. 

\bigskip

\noindent
Moving from on-shell amplitudes to off-shell aspects, manifestations of the double copy have been found for certain classes of exact classical solutions of GR and YM, being referred to as the \emph{Kerr-Schild double copy}. Initially established as a map from a stationary charge to the Schwarzschild metric \cite{Monteiro:2014cda}, this classical double copy was soon afterwards extended to more general stationary and even time-dependent solutions \cite{Luna:2015paa, Goldberger:2016iau, Ridgway:2015fdl, Luna:2016due, Carrillo-Gonzalez:2017iyj, Adamo:2017nia, Bahjat-Abbas:2017htu}.  
This construction hinges on the space-time metric admitting so called \emph{Kerr-Schild coordinates} (see e.g. \cite{Stephani:2003tm}), leading to the special property that the non-linearities of the Einstein field equations, and consequently the non-linearities of the YM equations, are completely absent. This means that the Kerr-Schild double copy essentially is a mapping between linear solutions of GR and YM.

The existence of these relations raises the question whether there exists an off-shell double copy formulation that takes into account (non-linear) off-shell information. Such a correspondence would be highly non-trivial, since off-shell information, in contrast to amplitudes, depends on the redundancies of the field-theoretic description (such as field basis and symmetry considerations). This redundancy already played an important role for the Kerr-Schild double copies, where the diffeomorphism symmetry was essential to construct the coordinate system in which one recognizes the Schwarzschild metric as a double copy of static gauge charge \cite{Monteiro:2014cda}.   

We will highlight exactly such an off-shell correspondence between the aforementioned triplet of effective scalar theories of NLSM, DBI and SG in {\bf section 2}. By picking a field basis for which the non-linear symmetries each contain the same type of terms, here chosen to be of the form $\delta \phi = \mathcal{O}(\phi^0) + \mathcal{O}(\phi^2)$, the field equations also take a very similar form, with each theory involving a distinct number of space-time derivatives and flavour structures. 

Given these similarities, we show that one can transform kinematic into flavour information (and vice-versa) by expanding the the scalar fields and the parameters of the non-linear symmetries in terms of \emph{auxiliary flavour coordinates} according to $\phi\rightarrow \phi_a \theta^a $, where $\theta^a$ is the auxiliary coordinate, and where the scalar field on the RHS only depends on the space-time coordinate. Under the assumption that the auxiliary flavour-coordinates are Grassmanian, the transformations of this type constitute invertible mappings between field equations and symmetries of the three Goldstone theories. The existence of these off-shell double copy relations imply that these theories really are different manifestations of the same underlying structure, expressed in different flavour and kinematic spaces. 

\bigskip

\noindent
The picture that emerges from the above considerations is a web of dualities discussed in the concluding {\bf section 4}: the different theories comprising the tensor-kinematic, colour, flavour and scalar-kinematic factors can be graphically represented as the \emph{tetrahedron} in figure \ref{fig:tetrahedron}. The triplet of scalar theories that are investigated in this paper, related by flavour-kinematics duality,  can be found at the front of the bottom level, thereby lying on the \emph{self-interaction face} - defined as those theories that retain non-trivial interactions even when restricted to a single species. We furthermore outline the relations to the other theories including colour.

\section{Non-linear symmetries and off-shell duality}\label{section:off-shell}

Our focus in this paper will be on the three scalar field theories that both allow for a double copy formulation and are determined by a non-linear (NL) symmetry. We will use the form of the latter to single out a field basis for the Goldstone modes in which the double copy is manifest, already at off-shell level (instead of for on-shell amplitudes). We will subsequently explain how this allows for double copy relations between these theories, corresponding to off-shell flavour-kinematics duality.

\subsection{A triplet of Goldstone theories}\label{section:2.1}

We will first focus on the formulation of the three theories, and adapt our field basis for all theories such that the NL symmetries have a similar structure, consisting of two types of terms: $\delta \phi = \mathcal{O}(\phi^0) + \mathcal{O}(\phi^2)$. The first term is a generalised shift term, that is independent of the scalar fields and only depends on the parameters and possibly space-time coordinates. In contrast, the second term is quadratic in the scalar fields, and furthermore depends on the parameters and possibly a space-time derivative. The first term is responsible for the Goldstone interpretation of the scalar fields and induces soft limits such as the (generalised) Adler zero \cite{Cheung:2014dqa,Cheung:2016drk}; the second term reflects the non-Abelian nature of these spontaneously broken symmetries\footnote{A similar structure can be identied in fermion EFTs with non-linear supersymmetry \cite{Kallosh}.} \cite{Brauner, Werkman}. Note that all theories are manifestly parity even in this formulation.

\bigskip

\noindent
The first theory of this form is the {\it non-linear \sm}, corresponding to the breaking of internal symmetries \cite{CWZ, CCWZ}. The corresponding Goldstone modes parametrise a symmetric coset $G/H$. We will focus on the case  
\begin{align}
  \frac{SO(M+N)}{SO(M) \times SO(N)} \,, 
\end{align}
or with isometry group $SO(M,N)$ instead, for the opposite sign choice between the two types of terms in the NL symmetry. 
The scalars are then fundamental representations of both $SO(M)$ and $SO(N)$, and will be denoted by\footnote{This may remind the reader of the formulation of GR inspired by double field theory, see e.g.~\cite{Hohm, Twofold}.} $\phi^{a \bar a}$ - or in matrix notation as $\phi$. Note that both flavour parts $a$ and $\bar a$ are independent and not necessarily of the same dimension ($M \neq N$). This coset structure can be realised by the NL symmetry transformation (in matrix notation)
 \begin{align}
     \delta \phi = c + \phi c^T \phi \,, \label{NLSM-transf}
 \end{align}
in terms of a constant matrix $c$. Including indices, this would correspond to $\delta \phi^{a \bar b} = c^{a\bar b}+\phi^{a\bar c}c^{d\bar{c}}\phi^{d\bar{b}}$. Note that we have suppressed a dimensionful scale in the second term on the RHS that sets the cut-off scale for the EFT, sometimes referred to as the pion decay constant $F$. 

Note that the above coset differs from the chiral NL \sm~that is often discussed in the literature, based on the symmetry breaking pattern $(G \times G) / G_{\rm diag}$ with e.g.~the pion case having $G_{\rm diag}=SU(N)$. Our reasons for focussing on the special orthogonal one instead will become clear as we outline the relations to the other Goldstone scalar field theories. Moreover, the cosets are not unrelated: upon identifying the two types of indices, $a = \bar a$ (which requires $M=N$), one can specialise to either the symmetric or anti-symmetric case with $\phi = \pm \phi^T$. The former corresponds to the coset $SL(N) / SO(N)$ while the second leads to the chiral one $(SO(N) \times SO(N)) / SO(N)_{\rm diag}$. As $SU(N/2)$ can be embedded in $SO(N)$ this contains the usual pion case. Moreover, the general $SO(M+N)$ case can be specialised to the $SO(M+1)$ case, which is relevant for e.g.~the composite Higgs model with $SO(5) / SO(4)$. See \cite{Low-soft} for a discussion of the $SO(M+1)$ case from the soft bootstrap perspective. 

Returning to the $SO(M+N)$ coset, the lowest order invariant for these Goldstone modes is the two-derivative NL \sm~Lagrangian, which in terms of the group element $g$ reads 
\begin{equation}
    \mathcal{L} =\frac{F^2}{4}\left[\partial g \partial g^{-1}\right],
\end{equation}
where $[ ... ] $ denotes a trace over flavour indices and $F$ is the pion decay constant, which we take to be one for legibility. One representation of an $SO(M+N)$ group element is given by
\begin{equation}
    g=\begin{pmatrix}
A & B\\
-B^{T} & C
\end{pmatrix}\, ,
\end{equation}
where the matrices are written in terms of the $M \times N$ Goldstone modes $\phi$ as
\begin{align}
    A=\frac{1-\phi \phi^T}{1+\phi \phi^T}\,, \qquad B=\frac{2}{1+\phi \phi^T} \phi\,, \qquad C= \frac{1-\phi^T \phi}{1+\phi^T \phi}\,,
\end{align}
which have the correct dimensions and properties to make up the $SO(M+N)$ matrix we need. This Lagrangian can be rewritten into
\begin{align}\label{NLSM_Lagrangian}
    \mathcal{L} = - \tfrac12 \left[ \frac{1}{1 + \phi \phi^T} \partial \phi \frac{1}{1 + \phi^T \phi} \partial \phi^T\right] \,,
\end{align}
which is the form we will adhere to in the following. Given the two-derivative nature of this theory, the corresponding field equations are naturally second order. After a number of simple manipulations that amount to solving for the Laplacian of the scalar field in terms of other quantities, these take the form
 \begin{align}
      \Box \phi = 2\sum_{n=1}^\infty (-1)^{n-1}  [(\partial \phi) \phi^T (\phi \phi^T)^{n-1} (\partial \phi) ]\,, \label{LNSM-field-eq}
 \end{align}
in $SO(M) \times SO(N)$ matrix notation.

\bigskip

\noindent
The second theory will involve {\it Dirac-Born-Infeld (DBI) scalars} that arise from space-time symmetry breaking\footnote{Interestingly, in this case the Goldstone theorem \cite{Goldstone1, Goldstone2} that associates a massless mode to every broken generator no longer applies, see e.g.~\cite{Ivanov, Low, Klein}.}. We will employ the multi-field generalisation \cite{multiDBI} that corresponds to the symmetry breaking pattern 
 \begin{align}
     \frac{ISO(D+N)}{SO(D) \times SO(N)} \,, 
 \end{align}
 where $D$ refers to the space-time dimension (and we are cavalier about its signature; one of the $D$ dimensions is actually time-like). Again, a different sign choice will change the isometry group to $ISO(D,N)$ instead. The spontaneous breaking of translations in the internal dimensions results in a number of scalar fields $\phi^a$ that transform in the fundamental representation of the internal symmetry $SO(N)$. The NL symmetry takes the form
 \begin{align}
     \delta \phi^a = c^a + c^a{}_\mu x^\mu + c^b{}_\mu \phi^b \partial^\mu \phi^a \,, \label{DBI-transf}
 \end{align}
and consists of a constant and linear shift (corresponding to translations and boosts) as well as a quadratic part (from the non-Abelian nature of boosts). 

The invariant Lagrangian can be written in terms of the induced metric $g_{\mu\nu} = \eta_{\mu\nu} + \partial_\mu \phi_a \partial_\nu \phi^a$. At lowest order in derivatives, this is given by the measure \cite{multiDBI},
 \begin{align}
      \mathcal{L} = 1 - \sqrt{g}  = - \tfrac{1}{2} \partial_{\mu} \phi \cdot \partial^{\mu} \phi + \tfrac{1}{4}\left(\partial_{\mu} \phi \cdot \partial_{\nu} \phi\right)\left(\partial^{\mu} \phi \cdot \partial^{\nu} \phi\right) - \tfrac{1}{8}\left(\partial_{\mu} \phi \cdot \partial^{\mu} \phi\right)^{2}+\ldots\,,
 \end{align}
where dots indicate flavour contractions. The field equations can be brought to the form
 \begin{align}
     \Box \phi^a = \sum_{n=1}^\infty (-1)^{n-1} [ (\partial \partial \phi^a) (\partial \phi \cdot \partial \phi)^{n} ] \,, \label{DBI-field-eq}
 \end{align}
in matrix notation for the space-time indices (and similar for the trace $[...]$).

\bigskip

\noindent
Finally, the {\it special Galileon theory} \cite{Cheung:2014dqa, sglagrangian} involves only a single Goldstone mode, with NL symmetry
 \begin{align}
     \delta \phi = c + c_\mu x^\mu + c_{\mu \nu} (x^\mu x^\nu + \partial^\mu \phi \partial^\nu \phi ) \,. \label{SG-transf}
 \end{align}
where again the first two parts are Abelian shift symmetries, and only the tensor transformation (with traceless parameter and a field-dependent, quadratic part) corresponds to the non-Abelian part. The latter corresponds to the coset 
 \begin{align}
     \frac{ISU(D)}{SO(D)} \,,
 \end{align}
(where $ISU(D)$ denotes the semi-direct product $\mathbb{R}^D \rtimes SU(D)$) while the former are central extensions thereof. The opposite sign choice in the NL symmetry modifies the special unitary group to the special linear group (i.e.~corresponds to a different real section of the complex group), and can be seen as Goldstone mode for affine coordinate transformations \cite{Roest}.

The invariant Lagrangian is given by a sum (with specific coefficients) of all Galileon terms with an even number of fields. The Lagrangian in $D$ dimensions reads \cite{sglagrangian}
\begin{equation}
    \mathcal{L}_{\mathrm{SG}}= -\frac{1}{2}  \sum_{n=1}^{\left\lfloor\frac{D+1}{2}\right\rfloor} \frac{(-1)^{n-1}}{(2 n-1) ! }(\partial \phi)^{2} \mathcal{L}_{2 n-2}^{T D}\,.
\end{equation}
The total derivative terms are given by
\begin{equation}
\mathcal{L}_{n}^{\mathrm{TD}}=\sum_{p}(-1)^{p} \eta^{\mu_{1} p\left(\nu_{1}\right)} \eta^{\mu_{2} p\left(\nu_{2}\right)} \cdots \eta^{\mu_{n} p\left(\nu_{n}\right)} \left(\Phi_{\mu_{1} \nu_{1}} \Phi_{\mu_{2} \nu_{2}} \cdots \Phi_{\mu_{n} \nu_{n}}\right)\,,
\end{equation}
where the sum is taken over all permutations of the indices $\nu$, with the sign of the permutation given by $(-1)^p$. The three leading terms explicitly read
\begin{align}
\mathcal{L}_{0}^{\mathrm{TD}} = 1 \,, \quad \mathcal{L}_{2}^{\mathrm{TD}}  =[\Phi]^{2}-\left[\Phi^{2}\right] \,, \quad 
\mathcal{L}_{4}^{\mathrm{TD}} &=[\Phi]^{4}-6\left[\Phi^{2}\right][\Phi]^{2}+8\left[\Phi^{3}\right][\Phi]+3\left[\Phi^{2}\right]^{2}-6\left[\Phi^{4}\right] \,,
\end{align}
where $[\Phi\cdots\Phi]$ denotes the trace over a product of matrices $\Phi \equiv \partial \partial \phi$ (all referring to space-time indices). In contrast to the previous two theories, the number of interaction terms in the Lagrangian is finite. As a consequence, the field equations can be written in a form with a finite number of interactions, which is strikingly different from the NL \sm~and the DBI theory. However, the SG field equations can also be brought to a similar form with an infinite sum of interactions,
 \begin{align}
     \Box \phi = \sum_{n=1}^\infty\frac{(-1)^{n-1}}{2n+1} [ \Phi^{2n+1} ] \,, \label{SG-field-eq}
 \end{align}
after isolating the Laplacian on the field on the left hand side.

\bigskip

\noindent
The above theories all contain a single scale that sets the magnitude of the interaction; we have not explicitly included this scale in the above, but it can be reinstated using dimensional analysis. Taking this scale to be an imaginary parameter effectively takes one from the above compact expressions to their non-compact versions. There are interesting constraints from UV considerations on which of these two constitute EFTs with viable UV completions; for single scalars, only the $ISO(1,D)$ version of DBI satisfies such positivity bounds, and similarly the $ISL(D)  =  \mathbb{R}^D \rtimes SL(D)$ version of the special Galileon, see e.g.~\cite{Dubovsky, Melville}. 

Finally, we would like to point out that there is a specific freedom in the construction of the NL \sm~that is absent for the other two theories. This can be seen at the off-shell level from two perspectives. The first one would be to notice that the NL \sm~is the only symmetry that is internal and has constant (i.e.~space-time independent) parameters. Due to this, one can implement this symmetry on arbitrary backgrounds, and moreover it allows for dynamical gravity. In other words, the construction of the NL \sm~carries over without modification after coupling it to gravity (see the appendix for details on this Lagrangian). Interestingly, this introduces a second coupling constant; in addition to the pion decay constant, we now also have Newton's constant (or the Planck mass) as a free parameter. 

A closely related perspective on this freedom is offered by the coset constructions. As the SG and the DBI theories have space-time symmetries, the corresponding cosets (of broken and unbroken symmetries) therefore necessarily include the Lorentz symmetry. In contrast, the NL \sm~can be seen as a product of two cosets, the one corresponding to the pion sector and the other forming the background; for flat geometries, the latter is simply Poincare over Lorentz as global symmetries:
 \begin{align}\label{symNLSMg}
  \frac{ISO(D)}{SO(D)} \times \frac{SO(M+N)}{SO(M) \times SO(N)} \,,
 \end{align}
with the coordinates $x^\mu$ and the scalar fields $\phi_{a \bar{ b}}$ corresponding to the broken generators. Upon gauging the algebra of space-time symmetries, one obtains dynamical gravity and non-linear (and non-Abelian) diffeomorphisms. The product of both cosets thus leads to the possibility to introduce a parameter for each coset. 

\subsection{Flavour-kinematics duality at the off-shell level}\label{section:2.2}

The similar natures of the non-linear transformations and the field equations suggest a relation between these theories at the off-shell level. Indeed one can go from theory to theory by replacing flavour with kinematic information, or vice-versa. 

\bigskip

\noindent
We will start with the non-linear tensor transformation of the SG theory, and note that it can be written in the following compact form
 \begin{align}
      \delta \phi = p + \tfrac12 \partial^\mu \phi \partial_{\mu \nu} p \partial^{\nu} \phi \,, \label{SG-starting}
 \end{align}
where one should interpret the parameter $p$ as a quadratic polynomial in the space-time coordinates $x^\mu$. In order to introduce flavour, one can augment these coordinates with a set of auxiliary coordinates $\theta^a$; these can be taken as an auxiliary construct, introduced to unify the three different theories. The parameter is now taken to be linear in the novel coordinate, $p = p_a \theta^a$, where in turn the $p_a$ are at most linear in space-time coordinates. The above transformation then takes the form (summing over both types of coordinates in \eqref{SG-starting})
\begin{align}
      \delta \phi = p + \partial^\mu \phi \partial_{\mu a} p \partial^{a} \phi  = p + \partial^\mu \phi \partial_\mu p_a \phi^a \,,
 \end{align}
where in the final expression we have similarly expanded the scalar field as $\phi = \phi_a \theta^a$, i.e.~as linear in the flavour coordinates. Note that the resulting components $\phi_a$ only have space-time dependence. In this way, the resulting transformation law is identical to that of DBI \eqref{DBI-transf} after expanding the above expression along the flavour coordinates\footnote{In a similar approach, colour information was systematically replaced by kinematic information in \cite{Cheung:2021zvb}, only at the level of currents instead of fields as outlined here.}.

Going one step further, one can also replace the remaining dependence on space-time coordinates with another flavour coordinate, $\bar \theta^{\bar a}$. These are taken as independent from the flavour coordinates $\theta^a$ (and in particular should not be read as (anti-)holomorphic coordinates). The parameter can then be taken bilinear in both flavour coordinates, $p = p_{a \bar b} \theta^a \bar \theta^{\bar b}$, with constant coefficients $p_{a \bar b}$. Similarly, we will take the fields to be bilinear in these, $\phi = \phi_{a \bar b} \theta^a \bar \theta^{\bar b}$. This results in the transformation law
\begin{align}
      \delta \phi = p + \partial^a \phi \partial_{a \bar b} p \partial^{\bar b} \phi  = p + \bar \theta^{\bar a} \phi_{a \bar a} p_{a \bar b} \phi^{b \bar b} \theta^b \,.
 \end{align}
Again, after expanding along the flavour coordinates, this expression is identical to the NL \sm~~transformation law \eqref{NLSM-transf}. This demonstrates the close relations between the different NL symmetries: the transformation laws are identical upon the appropriate identification of flavour dependences.

\bigskip

\noindent
This discussion has a parallel for the field equations - with an interesting twist moreover. We will consider the effect of the flavour Ansatze outlined above. Taking $\phi$ to be purely space-time dependent corresponds to the SG field equation \eqref{SG-field-eq}. When instead taking it linear in a flavour coordinate, the LHS retains this linearity and hence is proportional to $\theta^a$. On the other hand, the non-linearities corresponding to interactions can yield higher-order expressions. 

For simplicity, let's first discuss the cubic term on the RHS of \eqref{SG-field-eq}. Summing over both space-time and flavour coordinates, this expression yields two types of terms: with either three pairs of space-time contractions, or two space-time and one flavour contraction (terms with more than one pair of flavour contractions vanish as they will involve multiple flavour derivatives on a single field, incompatible with the flavour Ansatz). Starting with the latter, it takes the form
 \begin{align}
   [ \Pi (\partial \phi \cdot \partial \phi) ] = \theta^a \partial_{\mu \nu} \phi_a \partial^\mu \phi_b \partial^\nu \phi^b \,, \label{cubic}
 \end{align}
where in the compact first expression, the trace and matrix multiplication are for space-time indices $\mu$, and the dot is for flavour indices $a$. Note that the $1/3$ coefficient of this cubic SG Galileon in \eqref{SG-field-eq} is cancelled by the three-fold choice to distribute the flavour indices over the trace. Stripping off the overall auxiliary flavour coordinate, the LHS and cubic interaction then exactly combine into the corresponding terms for the multi-DBI field equation \eqref{DBI-field-eq}. The same also works for higher-order terms.

So far the discussion is completely analogous to that of the NL symmetries. However, in contrast to that case, the field equations also generate higher-order terms. For the cubic term, it takes the form $[\Pi^3]$ and hence is cubic in flavour coordinates. There will be similar contributions from the quintic and higher-order terms in the field equations that also generate a $\theta^3$ term. We therefore conclude that the SG field equation is, under the flavour Ansatz, translated into a set of conditions on the multi-DBI scalar fields. One of these is exactly the DBI field equation, with contributions such as \eqref{cubic}. Others are higher-order, such as the cubic one - which is cubic in flavour coordinates, and cubic plus higher-order in fields. 

Given the exact mapping between the transformation laws and the distinct dependences on the flavour coordinates, these different equations must   be separately invariant. Indeed the lowest-order one is identical to the DBI field equation, and the higher-order ones must be analogous in that they are invariant conditions. In order to get rid of these, one can take different stances. One would be to explicitly truncate the field equation at order $\theta$, thus retaining the lowest-order contribution. Another would be to take the auxiliary flavour dimensions to be Grassmannian; when contracted with the trace, the anti-symmetry then kills this term. 

\bigskip

\noindent
Analogous considerations apply to the transition from DBI to the NL $\sigma-$model. Instead, we now take $\phi$ to be bilinear in the flavour coordinate $\phi^a = \phi^{a\bar{b}}\bar{\theta}_{\bar{b}}$. 

For simplicity, let's consider cubic RHS of the DBI field equation \eqref{DBI-field-eq}. The trace over two space-time matrices generates two types of terms: with either zero or one flavour contraction (note that the term involving two flavour contractions again vanishes due to linearity in the flavour coordinate). Like before, the former is proportional to $\bar{\theta}^3$ and therefore vanishes under the assumption that the flavour dimensions are Grassmannian. The latter takes the form  
\begin{equation}
    2\bar{\theta}_{\bar{d}}\partial_\mu \phi^{a\bar{b}}\phi^{c\bar{b}}\partial^\mu \phi^{c\bar{d}} \,,
\end{equation}
where the factor two follows from the two-fold possibility to distribute the flavour indices within the trace. Stripping off the auxiliary flavour coordinate, the LHS and cubic interaction exactly coincide with the cubic part of the NL $\sigma-$model field equation \eqref{LNSM-field-eq}. Again, this also works for the higher-order interaction terms, where the two-fold choice remains because the flavour contraction needs to be on one of the partial derivatives on the two-derivative factor and on all contractions not involving it.  

\bigskip
\noindent
The above identifications lead to mappings of the three different theories with identical coupling constants (which we have not explicitly included for the moment). By allowing for numerical coefficients in the mapping, e.g.~$\phi = (M_{\rm SG} / M_{\rm DBI} ) \phi_a \theta^a$, one can also introduce arbitrary ratios. This will always map compact onto compact cosets, however. Finally, the flavour-kinematics duality as outlined here, i.e.~at the off-shell level, does not imply any restrictions or identifications between the two coupling constants of the NLSM; we will see in what follows that this will be different for the on-shell story.

\section{Scattering amplitudes and on-shell duality}\label{section:on-shell}

We now turn to the scattering amplitudes of the triplet of theories that we focus on. We will first discuss the BCJ formulation in terms of different kinematic numerators, and subsequently outline the relations between these numerators.

\subsection{The BCJ formulation}

\begin{figure}[t]
\centering
    \includegraphics[width=0.65\textwidth]{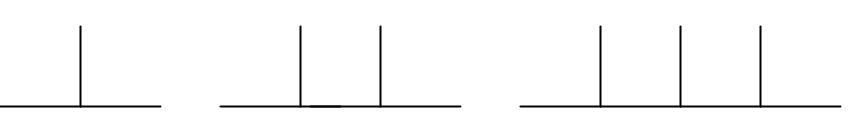}
    \includegraphics[width=0.65\textwidth]{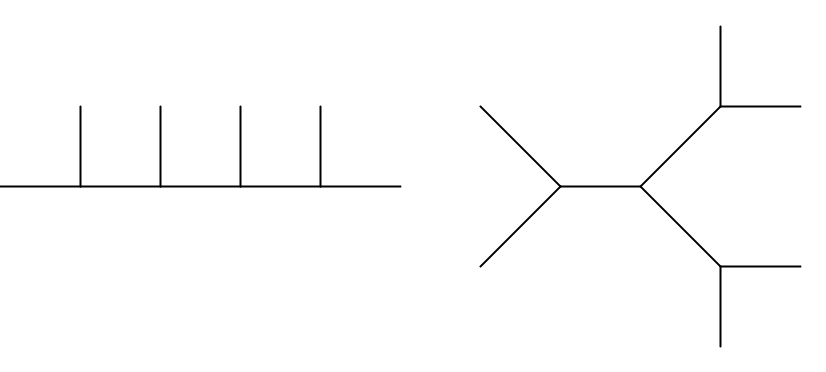}
  \caption{\it The trivalent diagrams relevant at lower-point scattering; all have half-ladder topology at three-, four- and five-points, while at six-points there are half-ladder and snow-flake topologies.}
  \label{fig:stu}
  \vspace{-0.25cm}
\end{figure}

In the BCJ formulation, amplitudes are generated by a sum over trivalent diagrams. As the first non-trivial example, the four-point amplitude can be written as
 \begin{align}
     A_4 = \sum_{\rm exchange} \frac{N_{ijkl} {\tilde N}_{ijkl}}{s_{ij}} \,,\label{4point-ampl}
 \end{align}
where the sum is over the three inequivalent exchange diagrams (see figure \ref{fig:stu}), corresponding to the permutations $(ijkl) = (1234, 2314, 3124)$ (i.e.~the $s$, $t$ and $u$ exchange channels), and the propagator structure in the denominator is $s_{ij} = (p_i + p_j)^2$. The reduction from in principle $4!$ permutations down to three is due to three $\mathbb{Z}_2$ symmetries: anti-symmetry in the first pair and in the last pair of indices, as well as order inversion. Note that these are all inequivalent trivalent diagrams when taking order invariance and anti-symmetry into account: using these, particle $4$ can always be placed at the final entry and there are three inequivalent options (particle $1,2,3$) for the penultimate entry.

More generally, this leads to $(2n-5)!!$ diagrams for general $n$-point amplitudes, leading to 15 and 105 contributions to five- and six-point amplitudes. The former consists of all halfladder permutations subject to the above $\mathbb{Z}_2$ symmetries (with $5! / 2^3 =15$ inequivalent possibilities). In the latter case, however, there are two inequivalent topologies, half-ladder and snow-flake diagrams, as is shown in figure \ref{fig:stu}.

There are $6!/2^3 = 90$ independent half-ladder permutations, with kinematic factors $N_{ijklmn}$. Next to these, there are $6! / (2^3 \cdot 3!)$ snow-flake permutations; there is a $3!$ reduction due to equivalence of the three legs, and $2^3$ due to the anti-symmetry in every leg. The kinematic factor for such diagrams is related via Jacobi identities and given by $N_{ijklmn} - N_{ijlkmn}$. The six-point amplitude then takes the form
 \begin{align}
     A_6 = \sum_{\rm half-ladder} \frac{N_{ijklmn} {\tilde N}_{ijklmn}}{s_{ij} s_{ijk} s_{mn}} + \sum_{\rm snow-flake} \frac{(N_{ijklmn} - N_{ijlkmn}) ({\tilde N}_{ijklmn} - {\tilde N}_{ijlkmn})}{s_{ij} s_{kl} s_{mn}}\,,
 \end{align}
in terms of the two numerators $N$ and $\tilde N$. Note the two different propagator structures in the denominators (where $s_{ijk} = (p_i + p_j + p_k)^2$).

\bigskip

\noindent
We now turn to the algebraic conditions on the numerators. In the BCJ formulation, the four-point amplitude factorises into kinematic numerators. These are required to satisfy the following symmetry conditions: 
 \begin{align}
     N_{ijkl} = - N_{jikl} \,, \quad N_{ijkl} = N_{lkji} \,, \quad N_{ijkl} + N_{jkil} + N_{kijl} = 0 \,. \label{4point-Jacobi}
 \end{align}
In addition to anti-symmetry in the first pair of indices and the order invariance, the third condition is often seen as a kinematic version of the Jacobi identity on structure constants.  

There is a natural generalisation of this story to higher $n$-point factors. These are subject to the following Jacobi-like identities, 
\begin{align}
    - N_{ijkl...} = N_{jikl...} = N_{k[ij]l...} = N_{l[[ij]k]...} = ... \label{Jac1}
\end{align}
involving multiple commutators for the first $n-1$ indices. In total there will be $n-2$ conditions of this form, generalising the two for the $4$-point function mentioned above. In addition, one can impose the order reverse condition 
 \begin{align}
     N_{ijklm...} = (-)^{n} N_{...mlkji} \,. \label{Jac2}
 \end{align}
These conditions translate into constraints on the possible representations that the factors can take. The corresponding Young tableaux are illustrated in figure \ref{fig:Young} and are given by the following:
\begin{itemize}
    \item For three-point factors, one can only have the anti-symmetric tensor. Its dimension (as element of the symmetric group) is $1$.
    \item At four-point, the unique representation is the window tensor with dimension $2$.
    \item At five-foint, we find the equal-arms hook tensor with dimension $6$.
    \item At six-point, we find three irreps with different Young tableaux. Their dimensions are $5$, $9$ and $10$, respectively, adding up to $24$.
\end{itemize}
Interestingly, these (ir)reducible representations have dimensions equal to $(n-2)!$ for every $n$-point factor. 

\bigskip

\noindent
The subsequent question is how to find specific representations that solve these conditions. For colour, these can be represented in terms of the structure constants of the group (where $A$ is the adjoint representation), and read
 \begin{align}
     N_{1234...} = f_{AB}{}^P f_{PC}{}^Q f_{QD}{}^R .... \,, \label{colour-commutator}
 \end{align}
involving commutators of the first $n-1$ indices. So for instance, at three and four-point, these are\footnote{In order not to clutter expressions, we are suppressing the adjoint indices on the LHS here; one should read this as e.g.~$(N_{123})_{ABC} = f_{ABC}$, with the understanding that $A$ refers to the adjoint rep of the first particle etc.}
 \begin{align}
     N_{123} = f_{ABC} \,, \quad 
     N_{1234} = f_{AB}{}^P f_{PCD} \,.
 \end{align}
Note that we assume the existence of a metric in the adjoint representation to lower the last index. These numerators naturally satisfy the Jacobi identities outlined above, and give rise to unique BCJ representations for colour.

\subsection{Flavour and kinematics numerators}

While the original BCJ paper addresses the factorisation of the scattering amplitudes of Yang-Mills and gravity in terms of colour numerators and kinematic numerators that involve momenta and polarisations, there have been many generalisations since. One of these is the identification of a numerator factor that solely depends on the Mandelstam variables, without polarisation. As a consequence, this describes the scattering of scalar particles, without additional structure (such as colour, flavour or spin). We first outline this possibility before considering generalisations that include flavour.

\begin{figure}[t!]
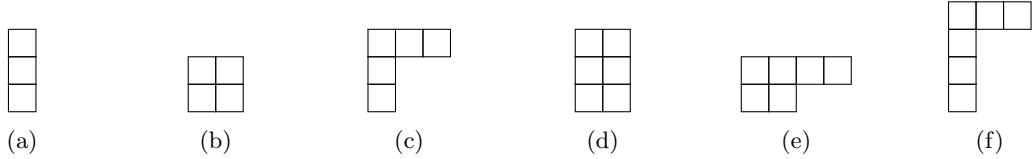

     \centering
     \begin{subfigure}[b]{0.15\textwidth}
         \centering
        \ytableausetup{boxsize=1em}
        \begin{ytableau}
         \  \\
         \   \\
         \ 
        \end{ytableau}   
        \subcaption{}
     \end{subfigure}
  \hspace{0cm}
     \begin{subfigure}[b]{0.15\textwidth}
         \centering
        \ytableausetup{boxsize=1em}
        \begin{ytableau}
 \ &\ \\
 \ &\  
\end{ytableau}   
\subcaption{}
     \end{subfigure}
  \hspace{0cm}
     \begin{subfigure}[b]{0.15\textwidth}
         \centering
        \ytableausetup{boxsize=1em}
    \begin{ytableau}
 \ &\ &\  \\
 \   \\
\ 
\end{ytableau}  
\subcaption{}
     \end{subfigure}     
  \hspace{0cm}
     \begin{subfigure}[b]{0.15\textwidth}
         \centering
        \ytableausetup{boxsize=1em}
        \begin{ytableau}
 \  &\ \\
 \ &\   \\
\ &\ 
\end{ytableau}   
\subcaption{}
     \end{subfigure}
  \hspace{0cm}
     \begin{subfigure}[b]{0.15\textwidth}
         \centering
        \ytableausetup{boxsize=1em}
        \begin{ytableau}
 \ &\ &\ &\ \\
 \ &\  
\end{ytableau}   
\subcaption{}
     \end{subfigure}
  \hspace{0cm}
     \begin{subfigure}[b]{0.15\textwidth}
 \centering
\begin{ytableau}
 \ &\ &\  \\
 \   \\
\ \\
\ 
\end{ytableau}  
\subcaption{}
\end{subfigure}
        \caption{\it Young tableaux for the BCJ numerators at a) three-, b) four-, c) five-, and d,e,f) six-point.} \label{fig:Young}
        \vspace{-0.25cm}
\end{figure}

\bigskip

\noindent
For scalar theories, the kinematic numerators can only depend on Mandelstam variables. The absence of colour structure implies that no anti-symmetric three-point numerators exist in the kinematic case, since all momentum contractions vanish under momentum conservation for massless particles. In this sense there is no kinematic analogue of structure constants. 

At four-point, one has to impose the Jacobi identities\footnote{An alternative to solving Jacobi identities was investigated in \cite{Roosmale}, by introducing the notion of numerator seeds. Following the rewriting of structure constants as a specific linear combination of traces of generators, their idea is to identify the (simpler) kinematic seeds that are the analogue of traces of generators. Subsequently, the full numerators are constructed by multiplying the seeds with a matrix $J$ encoding the Jacobi-like identities.} \eqref{4point-Jacobi}. We will be interested in factors that are quadratic in the Mandelstam variables\footnote{There is also a linear solution, with $N_{ijkl} \sim (s_{jk} - s_{ik} )$, that will however have a more natural interpretation when including flavour later on. It can also be used as a building block to generate quadratic and higher solutions along the lines of \cite{Rodina}. Monomials of higher order than quadratic could correspond to e.g.~higher-derivative corrections to the theories that we consider; however, we will restrict ourselves to the first non-trivial possibilities. \label{footnote:linear}}, in order to have the correct number of derivatives to connect to the special Galileon. The most general solution then reads
 \begin{align}\label{SKnumfourpoint}
     N_{ijkl} = \lambda_4 s_{ij} ( s_{jk} - s_{ik} ) \,,
 \end{align}
for the $s$-channel contribution. The resulting amplitude is 
 \begin{align}
     A_4 = -9\lambda_4{}^2 s_{12}s_{23}s_{13}
 \end{align}
This four-point amplitude indeed coincides with the special Galileon interaction of \eqref{SG-field-eq}. Note that, despite the propagators in the BCJ formulation of the amplitude \eqref{4point-ampl}, there are no poles over any Mandelstam variables in the result. This coincides of course with the statement that the theory under consideration has no three-point interactions that the four-point amplitude could factorise into.

The next case to consider are the 5-point factors. In this case, there are four constraints on the factors: in addition to the anti-symmetry and the order reversion ones, there are two Jacobi-like constraints involving commutators. We have checked that these constraints have no solutions below cubic order in Mandelstam. At cubic order, there is a single parameter. However, it turns out that the full amplitude vanishes in this case; the parameter should therefore correspond to a generalised gauge transformation. One can see this in the following way. 

Suppose the numerator would be (or would contain terms) of the following form:
 \begin{align}
     N_{ijklm} \sim s_{ij} G_{ijklm} - s_{lm} G_{mlkji} \,.
 \end{align}
When tensored with an arbitrary other numerator $\tilde N$, the amplitude receives a contribution
 \begin{align}
     \sum \frac{{\tilde N}_{ijklm} G_{ijklm}}{s_{lm}} - \frac{{\tilde N}_{ijklm} G_{mlkji}}{s_{ij}} \,.
 \end{align}
Provided the gauge parameters $G_{ijklm}$ are fully anti-symmetric in the first three parameters $(ijk)$, these 30 terms nicely combine into 10 triplets, where each triplet shares a common denominator. For instance, there will be terms proportional to
 \begin{align}
     \frac{({\tilde N}_{ijklm} + {\tilde N}_{jkilm} + {\tilde N}_{kijlm}) G_{ijklm}}{s_{ij}} \,,
 \end{align}
and similar for the other denominators. Of course, these terms nicely combine into a Jacobi identity and therefore cancel. Note that this is independent of the specific form of the second numerator ${\tilde N}$; it only requires that these satisfy the Jacobi identities. It turns out that the most general 5-point factor is exactly of this form, which gauge parameter given by 
 \begin{align}
    G_{ijklm} = ( s_{il} s_{jm} - (l \leftrightarrow m)) + (cyclic) \,,
 \end{align}
where the two cyclic terms refer to cyclicity in $(ijk)$. Therefore, the most general kinematic 5-point numerator is a gauge transformation.

At the six-point level, we will be interested in quartic factors in Mandelstam variables in order to connect to e.g.~the special Galileon\footnote{There are also quadratic and cubic solutions, that however are again naturally interpreted in the context of flavour.\label{footnote:quadratic}}. Solving the Jacobi identities \eqref{Jac1} and \eqref{Jac2} leaves one with 23 free parameters. Additionally, we have to impose factorisation into even amplitudes: the resulting amplitude may not have any poles over single Mandelstam variables, as this would correspond to splitting into a three- and a five-point vertex. This constraint further reduces the free parameters to only six\footnote{Operationally, we have first required the correct factorisation for amplitudes with one colour factor, and subsequently checked that kinematic $\times$ kinematic amplitudes also factorise correctly.}. When calculating the amplitudes, one finds that these only depend on a single linear combination of these coefficients. We therefore find five gauge parameters and one physical one.

Motivated by the overall $s_{ij}$ dependences of the 4- and 5-point kinematic factors, we consider a similar Ansatz at 6-point:
 \begin{align}
     N_{ijklmn} = s_{ij} P_{ijklmn} + \rm{order~reversed} \,. \label{kin-6point1}
 \end{align}
It turns out this Ansatz contains one gauge parameter and the physical one. The latter corresponds to the explicit expression
\begin{align}
       P_{ijklmn} = &-s_{mn} \big(s_{jk} \big(-4 s_{in}+4 \big(s_{jl}+s_{kl}+s_{km}\big)+5 s_{lm}\big)+\notag \\
    &+s_{ik} \big(-4 s_{in}+4 \big(s_{jl}+s_{kl}\big)+s_{jk}+9 s_{lm}\big)+5 s_{ij} s_{ik}+4 s_{ij}^2+s_{ik}^2\big)+\notag\\
    &+4 \big(s_{ij}+s_{ik}+s_{jk}\big) \big(\big(s_{ik}+s_{jk}\big) \big(-s_{in}+s_{jk}+s_{jl}+s_{kl}\big)+s_{jk} s_{km}\big)\notag\\
    &-4 s_{lm} \big(s_{ik} \big(s_{jl}+s_{kl}\big)+s_{jk} s_{jl}\big)+s_{mn}^2 \big(4 s_{ij}+5 s_{ik}\big)\, ,\label{kin-6point2}
 \end{align}
up to a generalised gauge transformation. The resulting amplitude agrees with the special Galileon 6-point interaction.

We expect that this structure of kinematic factors continues at higher points as well, with the relevant non-trivial solution to the Jacobi's coming in at order $n-2$ in Mandelstam variables. At odd points, these should be pure gauge (as the NLSM has only even amplitudes), while the even-point factors will be unique up to gauge transformations\footnote{Explicit expressions at six- and higher-point can be found in \cite{Du:2016tbc}; however they are constructed in a basis where the order reversion symmetry is not imposed.}.

\bigskip

\noindent
An important aspect of the discussion above is that there are no solutions to the Jacobi identities below a specific order in momenta. However, this assumes a structureless scalar field; new possibilities open up when augmenting the scalar field with additional structure. One example of this is to consider scalars in the biadjoint representation; this leads to the colour factors, starting with the structure constants $f_{ABC}$ as the three-point factor (where $A$ is the adjoint). Instead, we will focus on the fundamental representation, e.g.~of the special orthogonal group. Note that this directly eliminates all odd-point factors, as we will only be using the invariant metric $\delta_{ab}$. The remaining even-point expressions will be referred to as flavour factors.

For the flavour factors at four-point, we find the first non-trivial solution\footnote{The same factors were constructed in \cite{Low1,Low2} with a different interpretation, namely as higher-derivative corrections to a specific NLSM. We will comment on this possibility in the concluding section.} at linear order in Mandelstam variables:
 \begin{align}\label{flavour:4pt_factor}
    F_{1234} = f_1 ( \delta_{ab} \delta_{cd} (s_{23} - s_{13}) -  (\delta_{ac} \delta_{bd} - \delta_{bc} \delta_{ad}) s_{12} ) + f_2 (\delta_{ab}\delta_{cd}+\delta_{ac}\delta_{db}+\delta_{ad}\delta_{bc})(s_{13}-s_{23}) \,.
 \end{align}
Our notation here is that the first particle has momentum $p_1$ and flavour index $a$ etc. Note that the interplay between kinematic and flavour allows one to solve the Jacobi's in different ways; the above expression is anti-symmetric overall in the exchange of 1 and 2, that arises from anti-symmetry in flavour and symmetry in kinematic or vice versa. One consequence is that, in contrast to the colour and kinematic case, we find a free parameter already at the four-point level. When restricting to a single flavour, both parameters collapse onto the same factor, that was already mentioned in footnote \ref{footnote:linear}, which should therefore really be seen as a special case of the expression above including flavour.

By constructing amplitudes, one finds that both parameters $f_1$ and $f_2$ are actually physical instead of pure gauge. For instance, one can build amplitudes from kinematic and flavour factors,
 \begin{align}
     A_4 = \sum_{\rm exchange} \frac{N_{ijkl} F_{ijkl}}{s_{ij}} \,.
 \end{align}
The resulting amplitude splits up into a number of sectors that have different flavour structures, akin to the partial amplitudes of colour structures. For instance, along the $\delta_{ab} \delta_{cd}$ part, one finds
 \begin{align}
     A_4^{ab,cd} =  -6f_1\lambda_4 (s_{23}s_{13}) -6 f_2\lambda_4 (s_{23}s_{13}-s_{12}^2) \,.
 \end{align}
Comparing to the explicit Lagrangian of section 2 and its Feynman rules, one concludes that the $f_1$ part of this expression corresponds to the multi-DBI theory, with the specific relation between the two quartic terms Tr$[ (\partial \phi^a \partial \phi^n)^2 ]$ and Tr$[ (\partial \phi^a \partial \phi^n)]^2$. The second parameter follows from only the second of these quartic types. We therefore conclude that for generic parameter values this theory has no clear Goldstone interpretation associated with spontaneous symmetry breaking; this is only true for $f_2 =0$, resulting in the BCJ formulation of multi-DBI.

\begin{figure}[t]
         \centering
         \includegraphics[height=2.5cm]{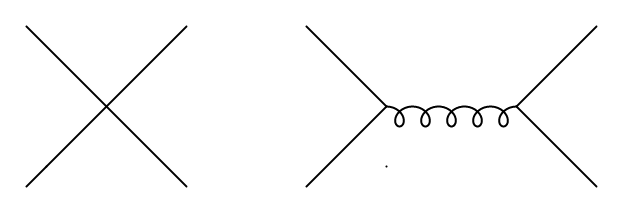}
        \caption{\it The two types of Feynman diagrams that contribute to the NLSM$_g$ four-point amplitude: contact interactions (left) and graviton exchange (right). Note that the straight lines are NLSM scalars and the curly line represents a (intermediate) graviton. }
        \label{fig:NLSMg4pt}
        \vspace{-0.25cm}
\end{figure}

As a second possibility, it is interesting to consider the product of two flavour numerators. In order to understand these amplitudes, it will be advantageous to introduce trace notation for the flavour structures generated by the $\delta$-functions, for instance 
 \begin{align}
  [AB] \equiv \delta_{ab} \delta_{\bar a \bar b} \,, \qquad [ABCD] \equiv \delta_{ab} \delta_{cd} \delta_{\bar{b}\bar{c}}\delta_{\bar{a}\bar{d}} \,,
 \end{align}
where $\delta$ and $\bar \delta$ generate the flavour structures of the two distinct numerators. In this notation, the amplitude contains two types of contributions:
 \begin{itemize}
     \item
     Firstly we have single-trace contributions, an example of which is given by
\begin{equation} 
\begin{aligned}
 \sim f_2^2\frac{(s_{12}^2-s_{23}s_{13})^2}{s_{12}s_{23}s_{13}}([ABCD]+[ADCB]) +(\mathrm{cyclic}) \, ,
    \end{aligned}
\end{equation}
where (cyclic) denotes cyclic permutation of three external legs, keeping one fixed. Note that the flavour structure of these terms, with a single trace, would perfectly correspond to the (expected) contributions from four-point operators in the NLSM; however, the kinematic structure includes (single) poles which would be incompatible with this. Therefore, we will henceforth set $f_2=0$. Note that this in retrospect justifies the interpretation of the four-point amplitude with one flavour and one kinematic factor as multi-DBI.

The only remaining single-trace contribution is given by
\begin{equation}
\begin{aligned}
    A_4^{\mathrm{contact}}=&-4f_1^2s_{13}([ABCD]+[ADCB]) +(\mathrm{cyclic}) \,. \\
    \end{aligned}
\end{equation}
These are indeed the amplitudes that would follow as contact diagrams from the Lagrangian \eqref{NLSM_Lagrangian}. 
\item
In addition to the single-trace terms, we have double-trace contributions of the form
\begin{equation}
     A_4^{\mathrm{exchange}}=-4f_1^2\frac{s_{23}s_{13}}{s_{12}}[AB][CD]+(\mathrm{cyclic}).
\end{equation}
Due to the pole structure, these cannot be contact terms; instead they correspond to graviton exchange between the four scalars (see figure \ref{fig:NLSMg4pt}); as the gravitons do not carry any flavour, this requires the flavours of the four particles to coincide pairwise\footnote{It would be interesting to investigate whether the poles introduced by $f_2$ have a similar interpretation as e.g.~gluon exchange; however, as we need to set $f_2$ to zero to get DBI and NLSM we will not pursue this option here.}. 
 \end{itemize}
Together, these single- and double-trace amplitudes arise from the scalar-sector of the $SO(M,N)$ non-compact NL \sm~minimally coupled to Einstein gravity, with the Lagrangian given by
\begin{equation}
    \mathcal{L}_{\mathrm{GR+NLSM}}=\sqrt{-g}\left(\frac{1}{2} M_{\rm Pl}^2 R -\frac{1}{2}  [ \frac{1}{1 - \phi \phi^T / F^2} \nabla^\mu \phi \frac{1}{1 - \phi^T \phi / F^2} \nabla_\mu \phi^T]\right)\,,
\end{equation}
where $M_{\rm Pl}$ is the (reduced) Planck mass and $F$ is the pion decay constant. Further details on the amplitudes of this theory are reviewed in the appendix. Given the relative strengths of the exchange and contact diagrams following from the BCJ prescription, we find that this corresponds to the non-compact NLSM coupled to gravity with the two coupling constants identified, $M_{\rm Pl} = F$, In contrast to the off-shell mapping outlined previously, the BCJ amplitudes therefore lead to a specific identification of the two parameters.

\bigskip

\noindent
At 6-point, there are 15 possible inequivalent flavour structures of the form $\delta_{\cdot \cdot} \delta_{\cdot \cdot} \delta_{\cdot \cdot}$. On the kinematic side, these would be multiplied by a quadratic expression in one of the 9 Mandelstam variables, leading to 45 different terms (and hence in total 675). Imposing the 6-point Jacobi identities constrain 642 of these parameters, leaving 33 parameters unfixed. These are split up amongst the 3 different irreps in the following way: the 5 contains 9 parameters, the 9 contains 15 parameters and the 10 contains 9 parameters. These then have to be further reduced to satisfy the factorisation constraints. Remarkably, this again leads to six parameters, similar to what was found for the kinematic factors. Moreover, all six parameters have components in all three irreps of figure \ref{fig:Young}.

Turning to the amplitudes, we find that these also depend on a single linear combination of these six parameters. We therefore again conclude that five correspond to a generalised gauge transformation, and there is a single physical parameter. The flavour factor $F_{123456}$ in general is given by a rather complicated expression. However, for a specific parameter (that is not equal to pure gauge) it follows from the flavour structure that multiplies the $s_{a,b}^2$ kinematics, which is given by
 \begin{align}
  & 28 \delta_{af} \delta_{be} \delta_{cd}-28 \delta_{ae} \delta_{bf} \delta_{cd}+11 \delta_{ab} \delta_{cd} \delta_{ef}+13 \delta_{af} \delta_{bd} \delta_{ce}-13 \delta_{ad} \delta_{bf} \delta_{ce} + \,, \notag \\
& -34 \delta_{ae} \delta_{bd} \delta_{cf}-5 \delta_{ad} \delta_{be} \delta_{cf}+25 \delta_{af} \delta_{bc} \delta_{de}-26 \delta_{ac} \delta_{bf} \delta_{de}+8 \delta_{ab} \delta_{cf} \delta_{de} + \,, \notag \\
 & -7 \delta_{ae} \delta_{bc} \delta_{df}+26 \delta_{ac} \delta_{be} \delta_{df}+13 \delta_{ab} \delta_{ce} \delta_{df}-9 \delta_{ad} \delta_{bc} \delta_{ef}+7 \delta_{ac} \delta_{bd} \delta_{ef} \,. \label{6pt-flavour}
 \end{align}
The other terms, that include the dependence on the other eight Mandelstam invariants, then follow from imposing both the Jacobis and the factorisation.

\subsection{Flavour-kinematics duality at the on-shell level}

\noindent
To complete the discussion of on-shell flavour-kinematics duality, finally we turn to the relations {\it between} the different factors. Analogous to the mapping of field equations and non-linear symmetries, one can also relate the different factors by a specific operation.

To illustrate this, we consider the four-point flavour factor
  \begin{equation}
   F_{1234} = \delta_{ab} \delta_{cd} (s_{23} - s_{13}) -  (\delta_{ac} \delta_{bd} - \delta_{bc} \delta_{ad}) s_{12}  \,.
  \end{equation}
Now replace the flavour information (i.e.~the $SO(N)$ delta-functions) with kinematic variables according to
\begin{equation}\label{fknummapping}
    \delta_{ab} \longmapsto 1 +\lambda \, s_{12} \,,
\end{equation}
where $\lambda$ is an arbitrary constant, and similar for all other flavour structures. We obtain an expression quadratic in $\lambda$, from which we isolate\footnote{Note that this mapping is not invertible, since we lose information by throwing away terms of highest order in $\lambda$.}  the part that is proportional to $\lambda$. Upon using four-point scattering identities (including e.g.~$s_{12} + s_{13} + s_{23} = 0$ and similar), the coefficient of this term turns out to be exactly the scalar-kinematic numerator \eqref{SKnumfourpoint},
 \begin{align}
     N_{1234} = s_{12}(s_{23}-s_{13})\,,
 \end{align}
after setting $\lambda = 1$. Note that the replacement \eqref{fknummapping} and subsequent restriction to terms linear in $\lambda$ implies that one goes from an amplitude that is quartic in derivatives to one that is sextic. This is the on-shell counterpart to the off-shell mapping discussed in section 2.2, which also introduces jumps by two derivatives at the four-point level.

The same mapping \eqref{fknummapping} has been verified at six-point scattering, where it relates the full six-point flavour amplitude (partly given by \eqref{6pt-flavour}) to the kinematic factor \eqref{kin-6point1} and \eqref{kin-6point2} after retaining the coefficient proportional to $\lambda^2$. Similarly, the five gauge parameters on the flavour and the scalar kinematic sides also map onto each other. We conjecture the same relation to extend to higher order as well.

\bigskip

While not the focus of this paper, it is interesting to note that there is a similar mapping from the tensor kinematic factor (relevant for GR and YM) onto the flavour factor. At four-point, the tensor factor is given by (see e.g.~\cite{Bern:2019prr}) 
\begin{equation}
 \begin{aligned}\label{TK4point}
    T_{1234} = 
-\big\{\big[\left(\varepsilon_{1} \cdot \varepsilon_{2}\right) p_{1}^{\mu}+&2\left(\varepsilon_{1} \cdot p_{2}\right) \varepsilon_{2}^{\mu}-(1 \leftrightarrow 2)\big]\big[\left(\varepsilon_{3} \cdot \varepsilon_{4}\right) p_{3 \mu}+2\left(\varepsilon_{3} \cdot p_{4}\right) \varepsilon_{4 \mu}-(3 \leftrightarrow 4)\big]\\
&+2s_{1,2}\big[\left(\varepsilon_{1} \cdot \varepsilon_{3}\right)\left(\varepsilon_{2} \cdot \varepsilon_{4}\right)-\left(\varepsilon_{1} \cdot \varepsilon_{4}\right)\left(\varepsilon_{2} \cdot \varepsilon_{3}\right)\big]\big\}\,.
 \end{aligned}
\end{equation}
To transform tensor-kinematic information into flavour-kinematics, we map\footnote{All odd-point tensor factors map onto vanishing flavour factors due to the odd numbers of polarisations and momenta.}
\begin{align}\label{transformations}
    \varepsilon_i\cdotp p_j &\longmapsto 0 \,,\quad 
    \varepsilon_i\cdotp \varepsilon_j  \longmapsto \delta_{ij}\,,
\end{align}
which results exactly in the flavour factors \eqref{flavour:4pt_factor}, with $f_1=1$ and $f_2=0$. Alternatively, one could take the starting point to be the polarization-stripped version of \eqref{TK4point}, which is given by 
\begin{equation}
\begin{gathered}
T_{1234}=\left[\left(-\frac{1}{2} \eta^{\alpha \beta} \eta^{\gamma \lambda} p_{1} \cdot p_{3}-\eta^{\gamma \lambda} p_{2}{ }^{\alpha} p_{3}{ }^{\beta}-\eta^{\alpha \beta} p_{1}{ }^{\lambda} p_{4}{ }^{\gamma}-2 \eta^{\beta \lambda} p_{2}{ }^{\alpha} p_{4}{ }^{\gamma}\right)\right. \\
\hspace{3cm}-(1 \leftrightarrow 2)-(3 \leftrightarrow 4)+(1 \leftrightarrow 2,3 \leftrightarrow 4) +\left(\eta^{\alpha \lambda} \eta^{\beta \gamma}-\eta^{\alpha \gamma} \eta^{\beta \lambda}\right) p_{1} \cdot p_{2}\bigg]\,,
\end{gathered}
\end{equation}
where we (analogous to flavour indices) associate the space-time index $\alpha$ with particle $1$, $\beta$ with particle $2$, and so forth. For the polarization-stripped numerator, the transformations \eqref{transformations} become 
\begin{align}\label{transform2}
    \eta^{\alpha\beta}\mapsto \delta^{ab}\,, 
\end{align}
while all terms involving two non-contracted momenta vanish. Note that the indices on the LHS of \eqref{transform2} are space-time indices, while the indices on the RHS represent flavour.

This relation between amplitudes spanned by tensor kinematics and flavour factors has a well-known counterpart at the Lagrangian level. The dimensional reduction of general relativity over $N$ dimensions leads to an $SL(N) / SO(N)$ coset, while when instead starting from the common sector $(g_{\mu\nu}, B_{\mu\nu}, \phi)$ in the higher dimensions this leads to an $SO(N,N)$ coset (see e.g.~\cite{odd-story}). In view of this, it should not be surprising that the BCJ factors of both theories, being the slightly generalised $SO(M,N)$ with $M \neq N$ and the common extension of GR with dilaton and two-form, are therefore also related.

In a similar vein, the mapping from tensor-kinematics to flavour and finally scalar-kinematics as outlined here is closely related to the operations \emph{dimensional reduction} (or \emph{compactify}) and \emph{generalized dimensional reduction} (or \emph{``compactify''}) of \cite{Cachazo}. In the CHY representation, the amplitudes are characterized by integrands that carry all information about the theory; each integrand consists of two building blocks containing polarizations and momenta. The (generalized) dimensional reduction procedure allows the polarizations of one of these building blocks to explore an internal space, thereby mapping e.g.~the gravity integrand to its Born-Infeld counterpart. Since gravity and Born-Infeld amplitudes can be constructed out of two tensor-kinematic numerators and a combination of a tensor-kinematic and a scalar-kinematic BCJ numerator respectively, the numerator mapping proposed here mimics this CHY operation at the level of the numerators. 

\section{Conclusion and outlook}\label{section:conclusion}

This paper deals with the interrelations between flavour and kinematic aspects of Goldstone theories. We have highlighted novel relations between three cases that display spontaneous symmetry breaking, ranging from internal (the pions of the $SO(M,N)$ NL \sm) to space-time symmetry (the $SO(N)$ multi-DBI scalars and the special Galileon). These theories therefore appear as different guises of the same underlying structure, which can be expressed in terms of flavour and/or kinematics. This flavour-kinematics duality results in the three Goldstone theories under study.

\begin{table}[t!]
\centering
\begin{tabular}{||c||c||c||} 
 \hline
  & \textbf{Non-linear symmetry} & \textbf{ Equation of motion} \\ [0.5ex] 
 \hline\hline
 \textbf{NLSM} & $\delta \phi =c +\phi c^T \phi$ & $\square \phi=\displaystyle\sum_{n=1}^{\infty}(-1)^{n-1} 2\left(\partial_{\mu} \phi\right) \phi^{T}\left(\phi \phi^{T}\right)^{n-1}\left(\partial^{\mu} \phi\right)$ \\
 \hline
 \textbf{DBI} & $\delta \phi^a =c^a +c^a{}_\mu x^\mu +c^b{}_\mu \phi^b \partial^\mu \phi^a $ & $\square \phi^{a}=\displaystyle\sum_{n=1}^{\infty}(-1)^{n-1}\left[\left(\partial \partial \phi^{a}\right)(\partial \phi \cdot \partial \phi)^{n}\right]$  \\
  \hline
  \textbf{SG} & $\delta \phi=c+c_{\mu} x^{\mu}+c_{\mu \nu}\left(x^{\mu} x^{\nu}+\partial^{\mu} \phi \partial^{\nu} \phi\right)$ &   $\square \phi=\displaystyle\sum_{n=1}^{\infty} \frac{(-1)^{n-1}}{2 n+1}\left[\Phi^{2 n+1}\right]$ \\
 \hline \hline
 \hline 
   & \textbf{Irrep} & \textbf{Factor at 4-point} \\ [0.5ex] 
 \hline\hline
 \textbf{Colour} & adjoint $A$ & $f_{AB}{}^E f_{CDE}$ \\ 
 \hline
 \textbf{Flavour} & fundamental $a$ & $\delta_{ab} \delta_{cd} (s_{23} - s_{13}) -  (\delta_{ac} \delta_{bd} - \delta_{bc} \delta_{ad}) s_{12} $ \\
  \hline
  \textbf{Kinematics} & singlet & $s_{12} ( s_{23} - s_{13} )$ \\  [1ex] 
 \hline
\end{tabular}  
\caption{\it The triplet of exceptional scalar EFTs with their non-linear symmetries and field equations, and the triplet of BCJ numerators into which they can be factorised. The off-shell mapping, as outlined in section \ref{section:2.2}, relates different theories while the on-shell mapping \eqref{fknummapping} relates different factors.}
\label{table:2}
\vspace{-0.25cm}
\end{table}

At the off-shell level, the field choice ambiguity can be fixed by requiring a common form of the NL symmetry transformation, with a constant and a quadratic part. In this basis, the field equations are seen to take closely related forms. These are summarised in the upper part of table \ref{table:2}. As outlined in section \ref{section:2.2}, the different fields, NL symmetries and field equations are mapped onto each other via an expansion in flavour coordinates $\theta^a$, with e.g.~the SG and DBI fields related as $\phi = \phi_a \theta^a$.

At the on-shell level, the amplitude of these theories can be built along the lines of BCJ factorisation from a range of building blocks. In addition to the well-known colour and kinematics factors, we have outlined how to construct flavour factors. These are subject to the same conditions, including group-theoretical (corresponding to the Jacobi-like identities) and physical (ensuring the correct amplitude factorisation) constraints. With the introduction of flavour, these factors employ the simplest $SO(N)$ representations, being scalar, fundamental and adjoint, see the lower part of table \ref{table:2}. As before, the reduction of flavour increases kinematics, as seen from the order in Mandelstam variables, and vice versa. Remarkably, the flavour and kinematics factor are found to be related by the simple substitution \eqref{fknummapping}; these therefore really correspond to the same structures expressed in different spaces.

\bigskip

\begin{figure}[t]
    \centering
    \includegraphics[scale=.74]{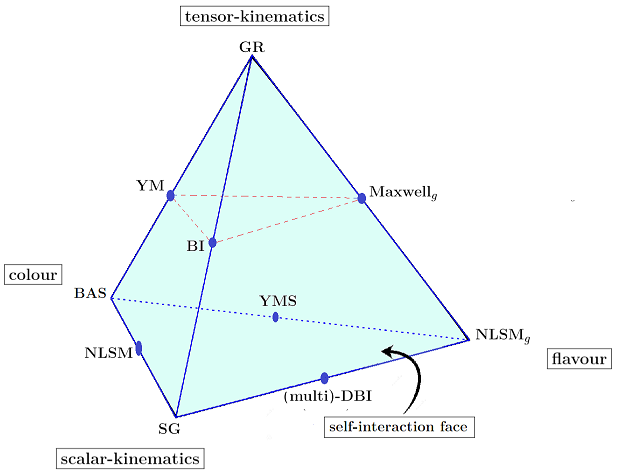}
    \caption{\it The tetrahedron spanned by the four different BCJ numerators that theories can factorise into. Taking all possible products results in the indicated web of dualities (including colour-(tensor-)kinematic and flavour-(scalar-)kinematic) between the different spin-2, -1, and -0 effective field theories. The right face of the tetrahedron corresponds to EFTs that retain interactions when restricted to a single species.}
    \label{fig:tetrahedron}
\vspace{-0.25cm}
\end{figure}

Which picture emerges from these different considerations? Employing the different combinations of factors, one can build a range of double copy theories. In addition to the three scalar factors describing colour, flavour and scalar kinematics, we will also include the possibility that was identified in the original BCJ proposal, including polarisation and thus tensor kinematics. Given these four factors, we find that the most direct graphical representation would be that of a tetrahedron, as illustrated in figure \ref{fig:tetrahedron}. The different nodes of this figure correspond to theories whose amplitudes are the products of the same type of factors, while one finds theories with mixed factors along the edges. 

Let us start with discussing the scalar theories at the bottom level of the diagram. At the front of the figure is the special Galileon theory, with two copies of scalar kinematics. When successively replacing scalar kinematics with flavour, one moves via the multi-DBI theory towards the $SO(M,N)$ NL \sm~minimally minimally coupled to gravity (and with the identification $F= M_{\rm Pl}$). When instead opting for colour instead of flavour, one encounters the NL \sm~without gravity, and finally the bi-adjoint scalar theory. The remaining scalar possibility has both a flavour and a colour factor, and corresponds to the scalar sector of the dimensional reduction of Yang-Mills (YMS)\footnote{Similar relations between the amplitudes of these sets of theories have been outlined from a different perspective in \cite{Cachazo,Cheung:2017ems }; our focus is on the BCJ formulation of these theories instead.}.

Moving up into the vertical direction, the original colour-kinematics duality corresponds to the edge linking the purely colour-based BAS theory and the tensor kinematics-based GR. Note that this is fully orthogonal to flavour-kinematics duality that we focus on and that interpolates between the SG and the gravity-coupled NL \sm. At the spin-1 intermediate level, one also encounters the Born-Infeld vector theory, as well as an $SO(N)$ multiplet of Maxwell vectors that only interact via gravity.

Remarkably, the $SO(M,N)$ NL \sm~appears twice in the above construction, with and without gravity-mediated interactions: the former consists of purely flavour, while the latter is the product of colour with scalar kinematics. This freedom was also encountered in the off-shell formulation of the theory, and corresponds to the introduction of an additional parameter arising from the direct product structure of the coset. Instead, from the BCJ factorisation, one either finds $M_{\rm Pl}$ infinite or equal to the pion decay constant. Perhaps one can consider combinations of these to interpolate to intermediate values of the Planck mass.

\bigskip

\noindent
A number of interesting questions appear naturally as a result of our findings. One of these concerns the self-interaction face on the right side of the tetrahedron. These theories retain non-trivial interactions when restricted to a single species (and hence no flavour or colour). As a consequence, these theories therefore will have non-linear responses to a source term, in contrast to e.g.~Yang-Mills theory. It would therefore be interesting to investigate the classical solutions of these theories and their possible mappings. Note that this would necessarily differ from the classical double copy as outlined in e.g.~\cite{Monteiro:2014cda}, as these map the linear Coulomb solution onto the Schwarzschild solution (effectively linearised when written in Kerr-Schild coordinates). Instead, the classical solutions on the self-interaction face would be non-linear, with a simple example provided by the Born-Infeld solution as a non-linear completion of the Coulomb solution \cite{Born:1934gh}.

Secondly, the flavour factors as identified in section 3 have already appeared in a different guise, namely as higher-derivative corrections to an $SO(M+1) / SO(M)$ coset. The addition of colour and flavour factors results in the interesting structure of extended DBI theory \cite{Low1, Low2}. Note that this combination is possible due to the special structure of the coset in the case of $N=1$, with colour and flavour both in the fundamental $SO(M)$ representation. It could be similarly interesting to further specialise to the case $M=1$, where flavour and scalar kinematics live in the same, trivial representation. The four-point factor for this theory are given by a linear combination of the expressions of footnote \ref{footnote:linear} and \eqref{SKnumfourpoint}. Perhaps this might give rise to a theory that includes DBI and SG as higher-derivative corrections to a free scalar field coupled to gravity.

Finally, it would be interesting to investigate whether the double copy that has been identified for on-shell amplitudes also extends to off-shell aspects such as correlators and wavefunctionals. This issue was addressed very recently \cite{Joyce}, where it was found that the most straightforward implementation of the double copy structure does not work (however, see also the analysis of \cite{Lipstein} employing cosmological scattering equations). Given the off-shell mapping outlined in section 2, it would be interesting to see whether this could be adapted for the gravity-coupled $SO(M,N)$ NL \sm, and whether this would allow for a mapping onto multi-DBI and SG.

\section*{Acknowledgments}

We are grateful to James Bonifacio, Kurt Hinterbichler, Austin Joyce, Ivan Kolar, Shubham Maheshwari, Anupam Mazumdar, Zhi-Zhen Wang for stimulating discussions on these and related topics. D.N. is supported by the Fundamentals of the Universe research program within the University of Groningen.

\appendix

\section{NLSM with GR} 

Here we derive the four-point amplitudes of the compact $SO(M+N)/ (SO(M) \times SO(N))$ NL \sm~ coupled to gravity. First, we will use the perturbation theory method as outlined in e.g. \cite{Luna:2016hge,Monteiro:2011pc}, after which we apply the \emph{Lehmann, Symanzik and Zimmermann (LSZ) formula} to extract amplitudes.

Recall from section \ref{section:on-shell} that the Lagrangian of the $SO(M+N)$ NL \sm~ minimally coupled to gravity, now including coupling constants, reads
\begin{equation}
    \mathcal{L}_{\mathrm{GR+NLSM}}=\sqrt{-g}\left(\frac{2}{\kappa^2}R -\frac{1}{2}  [ \frac{1}{1 + \frac{\phi \phi^T}{F^2}} \nabla^\mu \phi \frac{1}{1 + \frac{\phi^T \phi}{F^2}} \nabla_\mu \phi^T]\right)\,,
\end{equation}
where $\kappa^2= 4 / M_{\rm Pl}^2$ in terms of the (reduced) Planck mass, and $F$ is the NL \sm~cut-off scale. 
Following e.g. \cite{Luna:2016hge}, we work with the so-called \textit{gothic graviton} $\mathfrak{h}^{\mu\nu}$, such that
\begin{equation}
\sqrt{-g} g^{\mu \nu}=\eta^{\mu \nu}-\kappa \mathfrak{h}^{\mu \nu}\,.
\end{equation}
Furthermore, we adopt the \emph{De Donder gauge}, with $\partial_{\mu} \mathfrak{h}^{\mu \nu}=0$. These choices lead to the particularly useful properties that the Einstein tensor is given by $G_{\mu\nu}=-\frac{\kappa}{2}\square \mathfrak{h}_{\mu\nu}$, and the curved space-time d'Alembertian, denoted by $\square_c \equiv g^{\mu\nu}\nabla_\mu\nabla_\nu$, reduces to $\square_{c} = g^{\mu\nu}\partial_\mu\partial_\nu $ \cite{Donoghue:2017pgk}. The field equation for the graviton and scalar field respectively read
 \begin{align}
 \begin{split}
G_{\mu\nu} =& \frac{\kappa^2}{4}( g_{\rho\mu}g_{\sigma\nu}- \tfrac{1}{2} g_{\mu\nu}g_{\rho\sigma} ) [ \frac{1}{1 + \frac{\phi \phi^T}{F^2}} \partial^\rho \phi \frac{1}{1 + \frac{\phi^T \phi}{F^2}} \partial^\sigma \phi^T]\,, \label{fieldeq}  \\ 
   \square_{c} \phi=& 2\sum_{n=1}^{\infty}(-1)^{n-1}\partial_\mu \phi \frac{\phi^T}{F^2} \left(\frac{\phi \phi^T}{F^2}\right)^{n-1} \partial^\mu \phi \,. 
   \end{split}
\end{align}
By expanding $\mathfrak{h}^{\mu\nu}$ and $\phi$ in their coupling constants, 
\begin{equation}\mathfrak{h}^{\mu\nu}=\mathfrak{h}^{(0)\mu\nu}+\kappa \mathfrak{h}^{(1)\mu\nu}+\kappa^2 \mathfrak{h}^{(2)\mu\nu}+\ldots\,, \quad \:\:\; \phi=\phi^{(0)}+\frac{\phi^{(1)}}{F^2}+\frac{\phi^{(2)}}{F^4}+\ldots\,,
\end{equation}
and substituting these expansions into the field equations \eqref{fieldeq},
we obtain a differential equation for each perturbative correction $\mathfrak{h}^{(k)\mu\nu}$ and $\phi^{(k) }$. The relevant equations for four-scalar scattering are given by
\begin{equation}
\begin{gathered}
    \square\mathfrak{h}^{(0) \mu \nu} = - \frac{\kappa}{2} \left(\partial^\mu \phi^{(0)} \partial^\nu \phi^{(0)T} - \frac{1}{2} \eta^{\mu\nu} \partial^\rho \phi^{(0)}\partial_\rho \phi^{(0)T}\right) \,,\\\
    \square \phi^{(1)}=\kappa\partial_{\mu} \partial_{\nu} \phi^{(0)} \mathfrak{h}^{(0) \mu \nu}+\frac{2}{F^2}\partial^\mu \phi^{(0)}  \phi^{(0)T}\partial_{\mu} \phi^{(0)} \,. 
\end{gathered}
\end{equation}
Fourier transforming the above to momentum space leads to
\begin{equation}\label{h1}
\mathfrak{h}^{(0)\mu\nu}(-p_1)=-\frac{1}{p_1^2}\int \dbar^4p_2 \dbar^4 p_3 \frac{\kappa}{2}\left\{  (p_{2}^\mu p_{3}^\nu) -\frac{1}{2} \eta^{\mu\nu}(p_2\cdotp p_3) \right\}[CD][\phi^{(0)c\bar{c}}(p_2)\phi^{(0)d\bar{d}}(p_3)]\,,\end{equation}
\begin{equation}\label{momspace2}\begin{aligned}
\phi^{(1)a\bar{a}}(-p_1)
=-\frac{1}{p_1^2}\int \dbar^4p_2\dbar^4p_3\dbar^4p_4&\bigg\{\frac{\kappa^2}{4}\bigg(\frac{s_{23}s_{24}}{2s_{12}}-\frac{1}{2}(p_2)^2\bigg)[AB][CD] \phi^{(0)b\bar{b}}(p_2)[\phi^{(0)c\bar{c}}(p_3)\phi^{(0)d\bar{d}}(p_4)]\\
& - \frac{s_{13}}{2F^2}([ABCD]+[ADCB])\phi^{(0)b\bar{b}}(p_2)[\phi^{(0)c\bar{c}}(p_3)\phi^{(0)d\bar{d}}(p_4)]\bigg\}\,,
\end{aligned}\end{equation}
where we have explicitly included the flavour indices and suppressed the momentum-conserving delta functions $\deltabar^{(4)}(p_1+\ldots +p_n)$. Additionally, the common short-hand notation 
\begin{equation}
    \dbar^4p\equiv \frac{d^4 p}{(2\pi)^4}\,,  \quad   \deltabar^{(4)}(p)\equiv (2\pi)^4\delta^{(4)}(p)\,,
\end{equation}
was employed for legibility.

Next, we note that the term proportional to $p_2{}^2$ in \eqref{momspace2} vanishes on-shell and use the LSZ formula (see e.g. \cite{Monteiro:2011pc} for similar calculations) in order to extract the four-scalar partial amplitude from $\phi^{(1)}$.\footnote{ The LSZ formula extracts $n$-point amplitudes from $n$-point connected correlation functions (here corresponding to the perturbative corrections), by functionally differentiating $n-1$ times with respect to the leading order correction, while amputating the off-shell leg.} The result in terms of $M_{\rm Pl}$ reads
\begin{equation}
\begin{aligned}
    A_4 =& \lim_{p_1^2 \rightarrow 0 } p_1^2 \frac{\delta^3 \phi^{(1)}(-p_1)}{\delta \phi^{(0)}(p_2) \delta \phi^{(0)}(p_3) \delta \phi^{(0)}(p_4)}\\=& - \frac{1}{2M_{\rm Pl}^2} \frac{s_{14}s_{13}}{s_{12}} [AB][CD] +\frac{s_{13}}{2F^2}([ABCD]+[ADCB]) +(\text{cyclic})
    \,,
    \end{aligned}
\end{equation}
where the first term corresponds to graviton exchange diagrams and the second to contact interactions (see figure \ref{fig:NLSMg4pt}). The structures of these amplitudes coincide with the graviton exchange amplitude in equation $(11)$ of \cite{Draper:2020bop} and the NLSM amplitude in equation $(1.4)$ of \cite{NLSMLagrangian}. Note that we have opposite signs in the above amplitude, in contrast to what we found in section 3; the latter therefore corresponds to a non-compact scalar manifold with non-linear $SO(M,N)$ symmetry.

\bibliographystyle{alpha}


\end{document}